\documentclass[preprint,aps,nofootinbib]{revtex4}
\usepackage{graphicx}
\input{epsf.sty}
\usepackage{epsfig}

\newcommand{ \centeron }[2]{{\setbox0=\hbox{#1}\setbox1=\hbox{#2}\ifdim
                             \wd1>\wd0\kern.5\wd1\kern-.5\wd0\fi \copy0
                             \kern-.5\wd0\kern-.5\wd1\copy1\ifdim\wd0>\wd1
                             \kern.5\wd0\kern-.5\wd1\fi}}
\newcommand{ \ltap }{\>\centeron{\raise.35ex\hbox{$<$}}
                     {\lower.65ex\hbox{$\sim$}}\>}
\newcommand{ \gtap }{\>\centeron{\raise.35ex\hbox{$>$}}
                     {\lower.65ex\hbox{$\sim$}}\>}

\newcommand{ \lsim }{\mathrel{\ltap}}

\newcommand{ \slashchar }[1]{\setbox0=\hbox{$#1$}   
   \dimen0=\wd0                                     
   \setbox1=\hbox{/} \dimen1=\wd1                   
   \ifdim\dimen0>\dimen1                            
      \rlap{\hbox to \dimen0{\hfil/\hfil}}          
      #1                                            
   \else                                            
      \rlap{\hbox to \dimen1{\hfil$#1$\hfil}}       
      /                                             
   \fi}                                             %

\newcommand{ \mpl      }{M_{\rm Pl}}
\newcommand{ \ms    }{M_S^{}}
\newcommand{ \mssq  }{M_S^2}
\newcommand{ \xw  }{x_{\rm w}}

\def\singleandabitspaced{\baselineskip=\normalbaselineskip\multiply
    \baselineskip by 150\divide\baselineskip by 100}

\def\singlespaced{\baselineskip=\normalbaselineskip}

\begin{document}

\singlespaced


\hfill$\vcenter{
\hbox{\bf MADPH-04-1391}
\hbox{\bf FERMILAB-PUB-04-318-T}
                 \hbox{\bf hep-ph/0411094}}$
\vskip 0.4cm

\title{TeV-Scale String Resonances at Hadron Colliders} 
\author{Piyabut Burikham$^1$\footnote{piyabut@physics.wisc.edu},
Terrance Figy$^1$\footnote{terrance@physics.wisc.edu},
Tao Han$^{1,2,3}$\footnote{than@physics.wisc.edu} }
\vspace*{0.5cm}
\affiliation{$^1$Department of Physics, University of Wisconsin,
        1150 University Avenue, Madison, WI 53706, USA\\
        $^2$Theoretical Physics Department, Fermi National Accelerator Laboratory, 
        P.O.Box 500, MS106, Batavia, IL 60510, USA\\ 
        $^3$Institute of Theoretical Physics, Academia Sinica, Beijing 100080, China } 

\date{\today}

\vspace*{2.0cm}

\vskip 0.5cm

\begin{abstract}
We construct tree-level four-particle open-string amplitudes relevant to dilepton and diphoton 
production at hadron colliders.  We expand the amplitudes into string resonance (SR) 
contributions and compare  the total cross-section through the first SR with the 
$Z^{\prime}$ search at the Tevatron. We establish a current lower bound 
based on the CDF Run I results on the string scale to be about $1.1-2.1$ TeV, 
and it can be improved to about $1.5-3$ TeV with 2 fb$^{-1}$.
At the LHC, we investigate the properties of 
signals induced by string resonances in dilepton and diphoton processes. We demonstrate 
the unique aspects of SR-induced signals distinguishable from other new physics,
such as the angular distributions and forward-backward asymmetry.  
A $95\%$ C.L. lower bound can be reached at the LHC for  $M_S>8.2-10$ 
TeV with an integrated luminosity of $300$ fb$^{-1}$.  We emphasize the
generic features and profound implications of the amplitude construction.
\end{abstract}

\maketitle


\newpage

\setcounter{page}{2}
\renewcommand{\thefootnote}{\arabic{footnote}}
\setcounter{footnote}{0}
\singleandabitspaced

\section{Introduction}
\label{introduction-sec}

String theory \cite{books} remains to be the leading candidate to incorporate  
gravity into a unified quantum framework  of the elementary particle interactions.
The string scale ($\ms$) is naturally close to the quantum gravity scale
$\mpl \approx 10^{19}$ GeV, or to a grand unification (GUT) scale
$M_{GUT} \approx 10^{17}$ GeV \cite{Dienes:1996du}. It has been argued
recently that the fundamental string scale can be much lower  \cite{Lykken:1996fj}. 
With the existence of large effective volume of extra  dimensions beyond four, 
the fundamental quantum gravity scale may be as low as a TeV.
This is thought to have  provided an alternative 
approach to the hierarchy problem  \cite{add,rs},
namely the large gap between the electroweak scale $\cal O$(100 GeV) 
and the Planck scale of $\mpl$. What is extremely interesting is that these
scenarios would lead to very rich phenomenology at low energies
in particle physics \cite{shiu,extrad,pes} 
and astroparticle physics \cite{astro,cor,jos} that may be
observable in the next generation of experiments.

One generic feature of string models is the appearance of string resonances (SR) 
in scattering of particles in the energy region above the string scale.  
The scattering amplitudes are of the form of the Veneziano amplitudes \cite{books,has,pes},
which may develop simple poles. In the $s$-channel, the poles occur at 
$\sqrt s=\sqrt n \ms\ (n=1,2,\ ...)$ with degeneracy for different  angular momentum states.
It has been argued \cite{shiu,pes} that the scattering involving gravitons (closed strings)
is perturbatively suppressed  by higher power of string coupling 
with respect to the open-string scatterings which therefore 
are the dominant phenomena at energies near and above the string scale.

In this paper, we consider the possibility of producing 
the string resonances of a TeV-scale mass and studying their properties 
at colliders. 
We adopt the simplest open-string model in the D-brane scenario \cite{has,pes}.
It is assumed that all standard model (SM)
particles are identified as open strings confined to a D3-brane  universe, 
while a graviton is a closed string propagating freely in the bulk.  
For a given string realization of the SM, one should be able to calculate the open-string 
scattering amplitudes, in particular the Chan-Paton factors  \cite{cp} 
that are determined
by the group structure of the particle representations and their interactions. Unfortunately,  
there is no fully satisfactory construction of the SM from string theory and 
we are thus led to parameterize our ignorance.  
We demand that our stringy amplitudes reproduce the SM amplitudes at low energies.
The zero-modes of the scattering amplitudes are all identified 
as the massless SM particles and no new exotic states of the zero-modes are present.
By taking Chan-Paton factors to be free parameters,  a non-trivial stringy extension of 
the SM amplitudes to a higher energy region is accomplished by a unique matching 
between stringy amplitudes and those of the SM at low energies.

In fact, this scheme has been exploited in some earlier works.  These include possible
low-energy effects from the string amplitudes on four-fermion interactions \cite{bhhm}, 
and searching for signals in cosmic neutrino interactions \cite{cor,jos}.
In this paper, we explore the search and detailed study of their properties
for these string resonances  at hadron colliders such as the Fermilab Tevatron 
and the CERN Large Hadron Collider (LHC). 
In the string models, we expect a series of resonances 
with a predicted mass relation $\sqrt{n}M_S\ (n=1,2,\ ...)$.  Moreover, 
the angular distributions of the SR signals in parton-parton c.m.~frame 
present distinctive  shapes  in dileptonic and diphotonic channels due to the
angular momentum decomposition.  Rather small forward-backward asymmetry
is another feature of the model. These are all very unique and 
remarkably specific in contrast to signals from other sources of new physics.
It is found that the LHC experiments may be sensitive to a string scale of $\ms\sim 8$ TeV.  

The rest of the paper is organized as follows.
We first construct tree-level open-string scattering amplitudes for the dileptonic and 
diphotonic production processes in Sec.~\ref{ampl}, which reproduce the SM amplitudes 
at low energies and extend to include string resonances. 
In Sec.~\ref{pwe},  string resonance approximation is discussed and each string resonance 
is expanded into partial waves to see their angular momentum states.
Using the $Z^{\prime}$ constraints at the Tevatron, 
lower bounds on the string scale are obtained 
in Sec.~IV.   The analysis at the LHC is carried out in Sec.~V.  
We summarize in Sec.~VI our results and  emphasize the
generic features and profound implications of the amplitude construction. 
The complete expressions for the scattering amplitudes and the decay widths are given in two appendices.

\section{Construction of open-string amplitudes}
\label{ampl}

The 4-point tree-level open-string amplitudes can be expressed generically \cite{books,has,pes}
\begin{eqnarray}
A_{string} = S(s,t)\ A_{1234}\ T_{1234}+S(t,u)\ A_{1324}\ T_{1324}+S(u,s)\ A_{1243}\ T_{1243} 
\label{eq:1}
\end{eqnarray}
where (1, 2, 3, 4) represents external massless particles with incoming momenta.  
$A_{ijkl}$ are kinematic parts for $SU(N)$ amplitudes \cite{man}, which are given 
in Appendix A.  The Mandelstam variables at parton level are denoted by $s,t$ and $u$.  
For physical process ($12\to 34$), the $s,t$ and $u$-channels are labeled by
(1,2), (1,4) and (1,3), respectively.  $T_{ijkl}$ are the Chan-Paton factors and 
in the usual construction,
\begin{equation}
T_{1234}=tr(\lambda_{1}\lambda_{2}\lambda_{3}\lambda_{4})+tr(\lambda_{4}
\lambda_{3}\lambda_{2}\lambda_{1}).
\end{equation}
Following Ref.~\cite{man}, we adopt the  normalization of 
$tr(\lambda_{a}\lambda_{b})=\delta_{ab}$. 
Since a complete string model construction for 
the electroweak interaction of the standard model is unavailable, 
we will assume that these Chan-Paton factors are free parameters and 
$T_{ijkl}$ is typically in range of $-4$ to $4$.
$S(s,t)$ is essentially  the Veneziano amplitude
\begin{eqnarray}
S(s,t) & = & \frac{\Gamma(1-\alpha's)\Gamma(1-\alpha't)}{\Gamma(1-\alpha's-\alpha't)}
\label{eq:1.5}
\end{eqnarray}
where the Regge slope $\alpha'= M_S^{-2}$, 
and the amplitude approaches unity as either $s/\mssq$ or $t/\mssq \rightarrow 0$.

Of special interests for this article are the  $2 \rightarrow 2$ processes that may lead to
clear experimental signatures at the Tevatron and LHC.
We thus concentrate on two clean channels: 
the Drell-Yan (DY) dilepton production ($\ell\bar \ell$)
and the diphoton production ($\gamma\gamma$), 
from $q\bar q$ annihilation and possibly gluon-gluon  fusion.  
In this section, we explicitly construct the string amplitudes for these production processes.

\subsection{Dilepton Production}

At hadron colliders, the $2\to 2$ dilepton production processes are 
$q\bar{q}, gg \to \ell \bar \ell$.  The tree-level process for $gg \to \ell \bar \ell$
is absent in the SM. 
In the massless limit of the fermions, we label their helicities by the
chirality $\alpha,\beta=L,R$.
For the process with initial state $q\bar{q}$, we have two cases depending on the helicity combination
of the final state leptons. 
The non-vanishing amplitudes are
those for $\alpha \neq \beta$. The external particle ordering is ($12\to 34$).

\vskip 2mm
\noindent
\underline{(A1). $q\bar q$ annihilation  
$q_{\alpha}\bar{q}_{\beta} \rightarrow \ell_{\alpha}\bar{\ell}_{\beta}:$}
\vskip 2mm
With the notation as in Appendix A,  this  process 
belongs to a type of  $f^\pm f^\mp f^\mp f^\pm$,  with $\pm$ denoting the  helicity 
 of the particle with respect to incoming momentum.  Our construction thus leads to the
 physical amplitude    
\begin{eqnarray}
A_{string}(q_{\alpha}\bar{q}_{\beta} \rightarrow \ell_{\alpha}\bar{\ell}_{\beta})
= ig^2 \left[ T_{1234}S(s,t)\frac{t}{s}+T_{1324}S(t,u)\frac{t}{u}+T_{1243}S(u,s)\frac{t^2}{us} \right].
\label{eq:2}
\end{eqnarray}
The corresponding standard model amplitude is via the electroweak interaction,
\begin{eqnarray}
A_{SM} & = & ig_{L}^2\frac{t}{s}F_{\alpha\alpha},
\end{eqnarray}
where the photon and $Z$ contributions are given by
\begin{equation}
F_{\alpha\beta}=2Q_{\ell}Q_{q}\xw+\frac{s}{s-m^2_{Z}}\ \frac{2g^{\ell}_{\alpha}g^q_{\beta}}{1-\xw}.
\end{equation}
Here 
$\xw=\sin^2\theta_W$ and the $SU(2)_L$ coupling $g_{L}=e/\sin{\theta_{W}}$. The neutral
current couplings are $g^f_{L}=T_{3f}-Q_{f}\xw,\  g^f_{R}=-Q_{f}\xw$.

The crucial assumption for our approach is to demand the
string expression Eq.~(\ref{eq:2})  to reproduce the standard model amplitude 
in the low-energy limit when $s/\mssq \rightarrow 0$. This can be achieved by
identifying the string coupling with the gauge coupling $g=g_{L}$, 
and matching the Chan-Paton factors $T_{ijkl}$ as
\begin{eqnarray}
T_{1243}  =  T_{1324}\equiv T;\quad
T_{1234}  =  T+F_{\alpha\alpha}.
\end{eqnarray}
We then obtain the full result
\begin{eqnarray}
 A_{string}(q_{\alpha}\bar{q}_{\beta} \rightarrow \ell_{\alpha}\bar{\ell}_{\beta})
&=& ig^2_L S(s,t)\frac{t}{s}F_{\alpha\alpha} + ig_L^2T\frac{t}{us}f(s,t,u),
\label{eq:3} \\
 f(s,t,u) &=& uS(s,t) + sS(t,u) + tS(u,s).
\label{eq:f}
\end{eqnarray}
For simplicity, we will take the Chan-Paton parameter $T$ to be positive
and $0\le T\le 4$.  Taking $T$ to be negative will not change our numerical results appreciably.

A few interesting features are worthwhile commenting. First, we see that the string amplitude
Eq.~(\ref{eq:3}) consists of two terms: one proportional to the SM result multiplied by a
Veneziano amplitude $S(s,t)$;  the other purely with string origin 
proportional to an unknown Chan-Paton parameter $T$. In the low-energy limit $s\ll \mssq$,
$ f(s,t,u) \to s + t + u = 0$, reproducing the SM result regardless of $T$. This implies that
$T$ cannot be determined unless one specifies the 
detailed embedding of the SM to some more generalized group structure 
in a string setup.
The seemingly disturbing fact is that one of the Chan-Paton factors $T_{1234}$ must be
made dependent upon the $Z$-pole, rather than pure gauge couplings. This reflects our
ignorance of treating the electroweak symmetry breaking in our approach. 

As for the other helicity combination 
$q_{\alpha}\bar{q}_{\beta} \rightarrow \ell_{\beta}\bar{\ell}_{\alpha}$, it
belongs to the class of  $f^\pm f^\mp f^\pm f^\mp$. 
We apply the same methods as stated above and find the crossing relation
 $t \leftrightarrow u$ and an index interchange in the $F$ factor,
\begin{eqnarray}
A_{string}(q_{\alpha}\bar{q}_{\beta} \rightarrow \ell_{\beta}\bar{\ell}_{\alpha})
= ig^2_L S(s,u)\frac{u}{s}F_{\beta\alpha} + ig^2_L T\frac{u}{ts}f(s,t,u).
\label{eq:4}
\end{eqnarray}
with $T\equiv T_{1234}=T_{1324}$.

\vskip 2mm
\noindent
\underline{(A2). Gluon fusion $g_{\alpha}g_{\beta}\to\ell_{\alpha}\bar{\ell}_{\beta}:$}

In our open-string model, there is the possibility of dilepton production via two initial state gluons. 
 This amplitude vanishes at tree-level in the standard model, but could 
be non-zero in the open-string model if the gluons and leptons belong to some larger 
gauge group in which the Chan-Paton trace is non-vanishing.
The amplitude 
belongs to a type of  $g^\pm g^\mp f^\mp f^\pm$ according to Appendix A.
With $T\equiv T_{1234}=T_{1324}=T_{1243}$, the result reads 
\begin{eqnarray}
A_{string}(g_{\alpha} g_{\beta}\to\ell_{\alpha} \bar{\ell}_{\beta} )  
=  ig_L^2T\frac{1}{s}\sqrt{\frac{t}{u}}\ f(s,t,u) ,
\label{eq:5}
\end{eqnarray}
where $T$ may be different for  each helicity combination of external particles.  In fact, 
there exists an intrinsic ambiguity for the string coupling identification since there
are both strong interaction and electroweak interaction involved simultaneously.
Coupling identification for this subprocess would not be determined 
without an explicit string model construction. 
This problem is beyond the scope of this article. 
To be conservative, we have identified the string coupling with the weak coupling $g_L^{}$.

For $g_{\alpha}g_{\beta}\to \ell_{\beta}\bar{\ell}_{\alpha}$, we have $t \leftrightarrow u$ 
of the above expression.

\subsection{Diphoton Production}

Another clean signal in addition to dilepton production  at hadron colliders  
is the diphoton final state.  
We therefore construct the string amplitudes for diphoton processes in  this section.
We again label the helicities by $\alpha,\ \beta$, and as in the dileptonic processses,
the non-vanishing amplitudes are those with $\alpha\neq \beta$.

\vskip 2mm
\noindent 
\underline{(B1). $q\bar q$ annihilation 
$q_{\alpha}\bar{q}_{\beta} \to  \gamma_{\alpha}\gamma_{\beta}:$ }

Using the kinematic amplitudes for fermions and gauge bosons $f^\mp f^\pm g^\pm g^\mp$
as given in Appendix A and the matching techniques between 
the string and SM amplitudes described in the previous section, 
we obtain the following open-string amplitudes for $T\equiv T_{1234}=T_{1243}$,
\begin{eqnarray}
A_{string}(q_{\alpha}\bar{q}_{\beta} \to \gamma_{\alpha}\gamma_{\beta})  =  
2ie^2Q_{q}^2\sqrt{\frac{t}{u}}S(t,u) + ie^2 T\frac{1}{s}\sqrt{\frac{t}{u}}\ f(s,t,u),
\label{eq:16}
\end{eqnarray}
which correctly reproduce SM amplitudes at low energies, given by the
first term.  For the other helicity combination 
$\gamma_{\beta}\gamma_{\alpha}$, the amplitude
can be obtained by $t \leftrightarrow u$.

\vskip 2mm
\noindent
\underline{(B2). Gluon fusion 
$g_{\alpha}g_{\beta} \rightarrow \gamma_{\alpha}\gamma_{\beta}:$}  

Identifying this process with $g^\pm g^\mp g^\mp g^\pm$, one has
\begin{eqnarray}
A_{string}(g_{\alpha}g_{\beta} \rightarrow \gamma_{\alpha}\gamma_{\beta})
=ie^2T\frac{t}{us}f(s,t,u).
\label{eq:17}
\end{eqnarray}
with $T\equiv T_{1234}=T_{1324}=T_{1243}$.  Note that this amplitude is of purely  stringy origin.
There exists the same ambiguity for the string coupling identification 
as in $gg\to \ell\bar \ell$.  To be conservative, 
we have matched the string coupling with the electromagnetic interactions. 

For the other helicity combination $\gamma_{\beta}\gamma_{\alpha}$, the amplitude
can be obtained by $t \leftrightarrow u$.

\section{String Resonances and Partial Waves Expansion}
\label{pwe}

The factor $\Gamma(1-s/\mssq)$ in the Veneziano amplitude develops simple  
poles at  $s=nM^2_{S}\ (n=1,2,3...)$, implying resonant states with masses
$\sqrt n\ms$.
 At energies near the string scale, string resonances thus become dominating.  
 One can perform a  resonant expansion,  
\begin{eqnarray}
S(s,t) & \approx & \sum_{n=1}^{\infty}\frac{t(\frac{t}{M^2_{S}}+1)...(\frac{t}{M^2_{S}}+n-1)}{(n-1)!(s-nM^2_{S})}. \label{eq:7.1} 
\end{eqnarray}
Thus, by neglecting $S(t,u)$ which does not contain $s$-channel poles,
\begin{eqnarray}
f(s,t,u) & = & uS(s,t)+sS(t,u)+tS(u,s)
\nonumber\\
         & \approx & 2\sum_{n=\mbox{odd}}^{\infty}\frac{ut(\frac{t}{M^2_{S}}+1)...(\frac{t}{M^2_{S}}+n-1)}{(n-1)!(s-nM^2_{S})}.
\label{eq:7}
\end{eqnarray}
It is a remarkable result that this purely stringy function
$f(s,t,u)$ has only odd-$n$ SRs due to the crossing symmetry between $t$ and $u$.  
It represents the stringy effects of spin-excitations along the string worldsheet, which are
suppressed at low energy.  
These are the generic features of stringy effects we wish to explore 
at the high energy experiments.

\subsection{String Resonances in Dileptonic and Diphotonic Amplitudes}

The open-string amplitude construction for Drell-Yan processes predicts the existence of exotic intermediate states such as leptoquarks in the $u$-channel and higher spin bosonic 
excitations in the $s$-channel as string resonances. 
Due to the limited c.m.~energy accessible at collider experiments, 
 we need to keep only the first few resonances.  
 Applying the general results of Eqs.~(\ref{eq:7.1}) and (\ref{eq:7}) 
 to the dilepton string amplitudes,  we obtain the amplitude formula for the 
 first two resonances, with $\theta$ defined as angle between initial quark and final 
 anti-lepton in the parton c.m.~frame,
\[ A_{SR}(q_\alpha\bar q_\beta)  \approx \left\{ \begin{array}{ll}
 ig_L^2\frac{(1-\cos{\theta})^2}{4} \left[\frac{s}{s-M^2_S}(F_{\alpha\alpha}+2T) 
+ \frac{s}{s-2M^2_S}F_{\alpha\alpha} \cos{\theta}  \right]\quad & {\rm for}\ \ell_\alpha\bar{\ell}_\beta \\ 
\\ 
 ig_L^2\frac{(1+\cos{\theta})^2}{4} \left[ \frac{s }{s-M^2_S} (F_{\beta\alpha}+2T) 
- \frac{s}{s-2M^2_S}F_{\beta\alpha} \cos{\theta} \right] & {\rm for}\ \ell_\beta\bar{\ell}_\alpha. 
\end{array}
\right. \]
The full amplitude then will appear as a sum 
\begin{eqnarray}
A \approx A_{SM} + A_{SR}.
\end{eqnarray}

A few remarks on the amplitudes are in order.
Firstly, even we set free Chan-Paton parameter $T$ to zero, there are 
 still contributions from string resonances.
 This can be seen from the Veneziano factor multiplying to the SM 
 term in the string formula.  Significant differences from the
 standard model cross sections can be expected if the string scale is accessible
 at future colliders.  Second, the amplitude for the first (odd-$n$) string resonance depends 
on the Chan-Paton parameter $T$, while the second (even-$n$) resonance does not.
 The even resonances are completely determined by the gauge factors $F$ in 
 the standard model.
    
In the string model, there is a possible contribution from gluon fusion to lepton pairs, as seen
in Eq.~(\ref{eq:5}). Near the string resonance, we have
\begin{eqnarray}
A_{SR}(g_\alpha g_\beta)  \approx 
 ig_L^2  T\  \frac{s}{s-M^2_S}\  \frac{1\mp \cos{\theta}}{2} \sin{\theta},
\end{eqnarray}
%
where the sign $``-"$ corresponds to 
$g_{\alpha}g_{\beta} \to \ell_{\alpha}\bar{\ell}_{\beta}$, and $``+"$ to
$ \ell_{\beta}\bar{\ell}_{\alpha}$ with $\alpha\neq\beta$.
There are only odd-$n$ string resonances from this gluon contribution.  
This is generic for any processes if the standard model amplitude vanishes at tree-level.
It is always proportional to the function $f(s,t,u)$ which vanishes in the low energy limit,
which  only has odd-$n$ resonances. As a comparison, for processes with 
 the non-vanishing amplitudes in standard model at tree-level, their open-string 
 amplitude  will most likely contain both odd- and even-$n$ SRs.
 
The only exception is when the stringy correction piece multiplying to the standard model amplitude is $S(t,u)$ which does not contain SR pole in the $s$-channel.  This occurs naturally when the zero-mode (SM) tree-level exchange is in $t$ or $u$ but not in the $s$ channel.  We can see from the list in Appendix A that $A_{1324}$, to be multiplied with $S(t,u)$ in the full amplitude expression, never contain $s$-channel pole.  This is consistent with the physical picture that SR is the spin excitations of the zero-mode intermediate state.  If the zero-mode (SM) intermediate state does not exist, then there will not exist SR interacting with the same gauge charges.  An example of this kind of processes is $q\bar{q}\to \gamma\gamma$ which we can see from Eq.~(\ref{eq:16}).    
   For diphoton production, there are thus only odd-$n$ string resonances.    
   The first SR ($n=1$) for both processes are
\begin{eqnarray}
A_{SR}( q_{\alpha}\bar q_{\beta} \rightarrow \gamma_{\alpha}\gamma_{\beta}) 
& = & ie^2 T\ \frac{s}{s-M^2_S}\ \frac{1-\cos{\theta}}{2} \sin{\theta},  \label{eq:18}\\
A_{SR}(g_{\alpha}g_{\beta} \rightarrow \gamma_{\alpha}\gamma_{\beta})
 & = & 2ie^2T\  \frac{s}{s-M^2_S} \ \frac{(1-\cos{\theta})^2}{4}
 .\label{eq:19}
\end{eqnarray}     
The expressions for opposite helicity combinations $(\gamma_{\beta}\gamma_{\alpha})$
 are given by $\theta \to \pi-\theta$.  Observe that SR coupling is proportional to $T$ which is completely undetermined. 
 We will include these $n=1$ resonances and ignore those of $n=3$ 
 in our LHC analysis for the diphoton signals.

\subsection{Partial Waves Expansion of String Resonances}
\label{part}

There is degeneracy of states with different angular momenta at each SR as can be seen from the dependence on different powers of $t$ for each $n$ in Eq.~(\ref{eq:7.1}).  
Generically, any amplitude $A(s,t)$  can be expanded in terms of the Wigner functions $d^j_{mm^{\prime}}(\cos{\theta})$ \cite{pdg} as
\begin{eqnarray}
A(s,t) & = & 16\pi \sum_{j=M}^{\infty}(2j+1)a_{j}(s)d^j_{mm^{\prime}}(\cos{\theta})
\end{eqnarray}
where $M=\max(|m|,|m^{\prime}|)$, and  $a_{j}(s)$ are the partial wave amplitudes
corresponding to a definite angular momentum state $j$.

For our purpose, we expand the SR amplitudes for each mass eigenstate of a given $n$
by the Wigner functions as in Table~\ref{table:SRamps}.

\begin{table}[h]
\caption{
\label{table:SRamps}}
\tabcolsep=.5cm  
\def\arraystretch{2.5}  
\def\dis{\displaystyle}  
\begin{tabular}{|ll|}
\hline
\underline{DY dilepton pairs}&\\[-.35cm]
$A^{n=1}_{SR}(q_\alpha\bar q_\beta \to \ell_\alpha\bar{\ell}_\beta)$  &
$ ig_L^2 (F_{\alpha\alpha}+2T)
\dis \sum_{j=1}^{2}\frac{s\ \alpha^j_1\ d^j_{1,-1}}{s-M^2_S+i\Gamma^j_1 M_S}$
 \\
$A^{n=1}_{SR}(q_\alpha\bar q_\beta \to  \ell_\beta\bar{\ell}_\alpha)$ &
$ ig_L^2 (F_{\beta\alpha}+2T)
\dis \sum_{j=1}^{2}\frac{s\ \alpha^j_1\ d^j_{1,1}}{s-M^2_S+i\Gamma^j_1 M_S} $
\\
$A^{n=2}_{SR}(q_\alpha\bar q_\beta \to  \ell_\alpha\bar{\ell}_\beta)$  &
$ ig_L^2 \ F_{\alpha\alpha}
\dis \sum_{j=1}^3 \frac{s\ \alpha^j_1\ d^j_{1,-1}}{s-2M^2_S+i\Gamma^j_2 \sqrt{2}M_S}$
\\
$A^{n=2}_{SR}(q_\alpha\bar q_\beta \to \ell_\beta\bar{\ell}_\alpha)$ &
$ ig_L^2\ F_{\beta\alpha}
\dis \sum_{j=1}^3 \frac{s\ \alpha^j_1\ d^j_{1,1}}{s-2M^2_S+i\Gamma^j_2 \sqrt{2}M_S} $
\\
 \hline
 $A^{n=1}_{SR}(g_{\alpha}g_{\beta} \to \ell_{\alpha}\bar{\ell}_{\beta},\ \ell_{\beta}\bar{\ell}_{\alpha})$ &
$\dis  ig_L^2T \frac{s\ d^2_{2,\mp1}}{s-M^2_S+i\Gamma_1 M_S} $
 \\[.5cm]
 \hline
 \underline{Diphoton final state}&\\[-.35cm]
$A^{n=1}_{SR}( q_{\alpha}\bar q_{\beta} \rightarrow \gamma_{\alpha}\gamma_{\beta},\
\gamma_{\beta}\gamma_{\alpha})$ &
$\dis ie^2 T\ \frac{s\ d^2_{2,\mp 1}}{s-M^2_S+i\Gamma_1M_S} $
 \\
 $A^{n=1}_{SR}(g_{\alpha}g_{\beta} \rightarrow \gamma_{\alpha}\gamma_{\beta},\
\gamma_{\beta}{\gamma}_{\alpha})$ &
$\dis 2ie^2T\  \frac{s\ d^2_{2,\mp 2} }{s-M^2_S+i\Gamma_1M_S}$
\\[.5cm]
\hline
\end{tabular}
\end{table}

It becomes clear that the different angular momentum states will lead to
very distinctive angular distributions of the final state leptons for the 
 SR signals and may serve as important indicators in exploring the
 resonance properties.  To regularize the poles, 
  the decay widths have been included.  
 The coefficients $\alpha^j_n$, decay widths $\Gamma^j_n$, and the 
 relevant Wigner functions are given in Appendix B. 

\section{ Bounds on the String Scale from the Tevatron}

At the Fermilab Tevatron, the clean channels of dileptons and diphotons have been 
actively searched for. The CDF collaboration has been searching for a $Z'$ gauge boson
in the dilepton channel and a lower bound $M_{Z'}>690$ GeV had been
set based on their Run I data \cite{cdf} for a neutral gauge boson with SM-like couplings.
Similar results were obtained by the D0 collaboration \cite{d0}. 
The non-existence of a signal put an upper bound on the production cross section
and can thus be translated to  stringent constraints on the string scale. 

Using CTEQ5L  parton distribution functions \cite{cteq} , we estimate the total cross-sections 
for the string resonance signatures at various string scales with $T=1-4$.  
Since there is degeneracy of state with different angular momenta at the same mass, 
we use partial wave expansion to split 
each SR pole.  We regulate the resonance pole by including the decay width of each angular momentum state separately.  
The detailed treatment for the width calculation is given in Appendix B.  
For instance, 
for $M_S=1$ TeV, $n=1$ and $T=1$, the widths of SR in the Drell-Yan process 
are 240 (48) GeV for $j=1\ (2)$, while the width of SR in $gg \rightarrow \ell\bar{\ell}$
is 19 GeV with the only $j=2$ state.
When we compare with Tevatron  data on their $Z^{\prime}$ search, 
we need only the first SR,  the lightest state (including the angular momentum degeneracy).  

\begin{figure}[tb]
\centering
\epsfxsize=4.3in
\hspace*{0in}
\epsffile{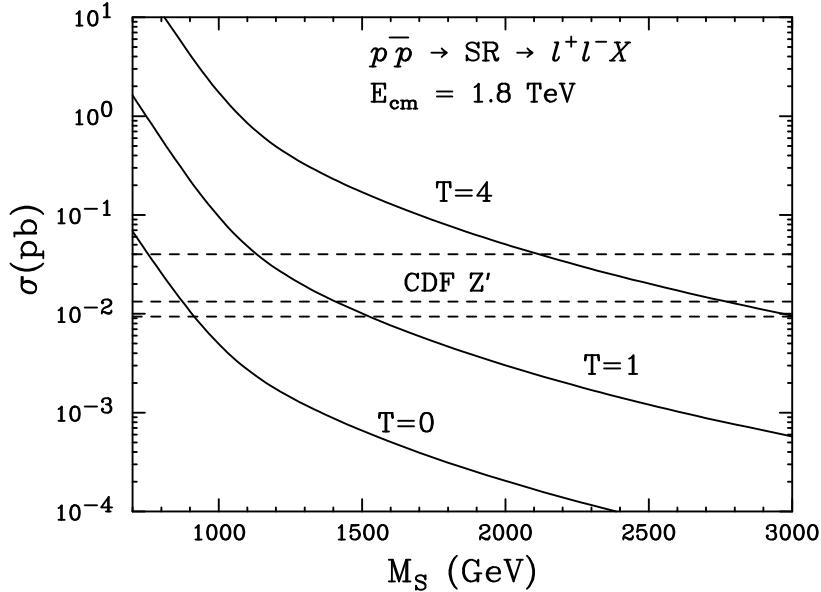}
\caption{Total cross section for the DY process ($\ell=e, \mu$) via the
SR versus its mass $\ms$, for different values  of $T=0-4$  (the solid curves). 
Detector acceptance cuts of Eq.~(\ref{tevcuts}) have been imposed.
The horizontal dashed lines show the $95\%$ C.L. upper bound 
on $\sigma(Z^{\prime})B(Z^{\prime}\to \ell\ell)$ for integrated  luminosities  
$110\mbox{ pb}^{-1}$, $1\mbox{ fb}^{-1}$ and $2\mbox{ fb}^{-1}$, respectively.  }
\label{tev}
\end{figure}

In Figure \ref{tev},  we present the total cross section for the DY process ($\ell=e,\mu$)
via the SR versus its mass $\ms$, for different values  of the Chan-Paton parameter
$T=0-4$ as shown by the solid curves.  Both contributions  from
$q\bar q$ and $gg$ are taken into account.
To extract the lower bound on the string scale, 
we have simulated the experimental acceptance cuts 
on the invariant mass of the lepton pair,
transverse momentum of the leptons, and their rapidity to be 
\begin{equation}
M(\ell\ell) > 50\  {\rm GeV},\quad  p_T(\ell) >18\ {\rm GeV},\quad |y_\ell | < 2.4.  
\label{tevcuts}
\end{equation}
We extrapolate CDF result \cite{cdf} of 110 pb$^{-1}$ 
on the $Z^{\prime}$ mass bound at $95\%$ C.L.
through dilepton production to a higher mass scale to obtain an upper bound on
the production cross section, 
as shown by the horizontal dashed lines, corresponding to different integrated luminosities, 
$110\mbox{ pb}^{-1}$, $1\mbox{ fb}^{-1}$ and $2\mbox{ fb}^{-1}$, respectively.
The intersections between the top horizontal line from the extrapolated data and  
the curves calculated for string resonances are located at $1.1-2.1$ TeV for $T=1-4$,
and thus yield the current lower bound on $\ms$.   
This gives a stronger  bound for the string scale than that based on a
contact interaction analysis \cite{bhhm}.
A bound obtained from the diphoton final state is weaker than that from 
the DY process, and we will not present it here.

In the near future with an integrated luminosity of
2 fb$^{-1}$ at the Tevatron, one should be able to extend
the search to $\ms\sim 1.5-3$ TeV for $T=1-4$, as indicated in Fig.~\ref{tev}.
It is interesting to note that even for $T=0$, one still has some sensitivity at
the Tevatron, reaching $\ms\sim 1$ TeV.

\section{String Resonances at the LHC}
\label{LHC-sec}

At the LHC, operating at $E_{cm}=14$ TeV with an expected luminosity of $300$ fb$^{-1}$, 
could produce a sufficiently large number of events induced  by SRs with masses of several TeV.
We will first present various aspects of dilepton and diphoton SR-induced signals in comparison 
with the expected SM backgrounds. Then we will proceed to set the lower bound
on the string scale if we do not see any SR-induced signals at the LHC.  
For illustration, we take a fixed string scale of $M_S=2$ TeV and $T=1$.  
All of the processes are calculated with the minimal acceptance cuts on the final state
particles of leptons and photons
\begin{equation}
\label{cuts}
  p_T >20\ {\rm GeV},\quad |y | < 2.4.  
\end{equation}
To be more realistic in generating the resonant structure, we smear the particle
energies according the electromagnetic calorimeter response with a Gaussian
distribution
\begin{equation}
{\Delta E\over E} = {5\%\over \sqrt{ E/{\rm GeV}} } \oplus 1\%.
\end{equation}

\begin{figure}
\centering
\epsfxsize=4.7in
\hspace*{0in}
\epsffile{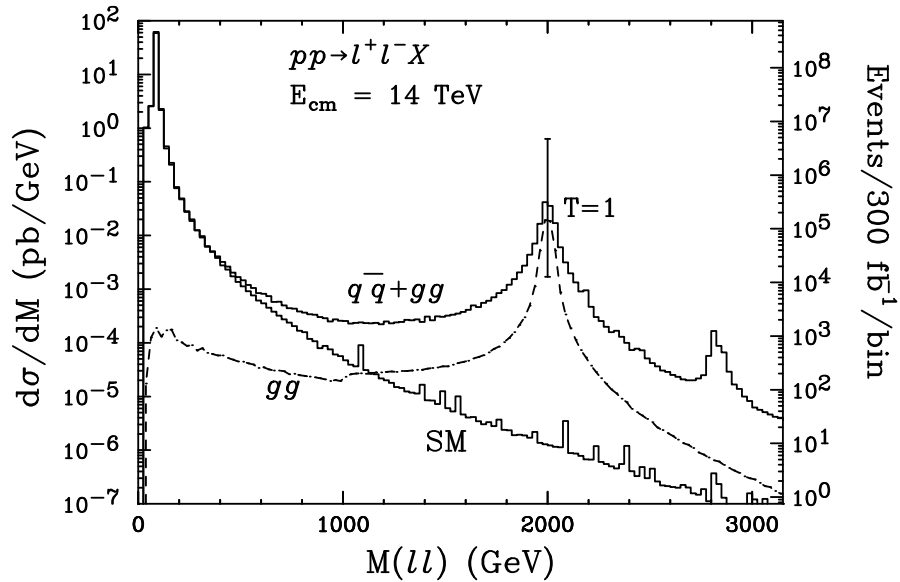}
\caption{Invariant mass distributions for DY dilepton production at the LHC, for the
continuum SM expectation and the SR contributions with $\ms=2$ TeV
and $T=1$: $q\bar q+gg$ (top curve) and $gg$ only (dashed). 
The vertical bar at the $n=1$ SR peak indicates the enhancement
for $T=4$. }
\label{mll}
\end{figure}

\subsection{The resonance signals}
In Figure \ref{mll}, we present the invariant mass distributions of the DY dileptons
for  the SM background expectation and the string resonances,  
 including both $q\bar q$ and $ gg$ contributions as labeled. 
At low energies, the stringy amplitudes reproduce SM results as expected. 
At higher energies, the resonant structure in the invariant
mass distribution can be very pronounced. 
The dilepton processes have both even- and odd-$n$ SRs, with masses  
$\ms,\ \sqrt{2}\ms$ for $n=1,2$.
 Recall that the second SR is  independent of the Chan-Paton parameter $T$, 
 in contrast to the first SR which is  dependent on $T$.  To illustrate this effect, 
 we have also depicted the peak height for the choice of $T=4$. Therefore, 
the number of events around the first SR (the cross section)
will determine the Chan-Paton parameter $T$,
while the number of events around the second SR will be predicted essentially
by the SM couplings.  
Moreover, the mass of the second string resonance is remarkably 
predicted to be $\sqrt{2}M_S$, fixed with respect to  the first resonance.  
These essential aspects of SR signals allow us to distinguish this unique model from other 
new physics. The scale on the right-hand side gives the number of events per bin
for an integrated luminosity of 300 fb$^{-1}$.

\begin{figure}
\centering
\epsfxsize=4.7in
\hspace*{0in}
\epsffile{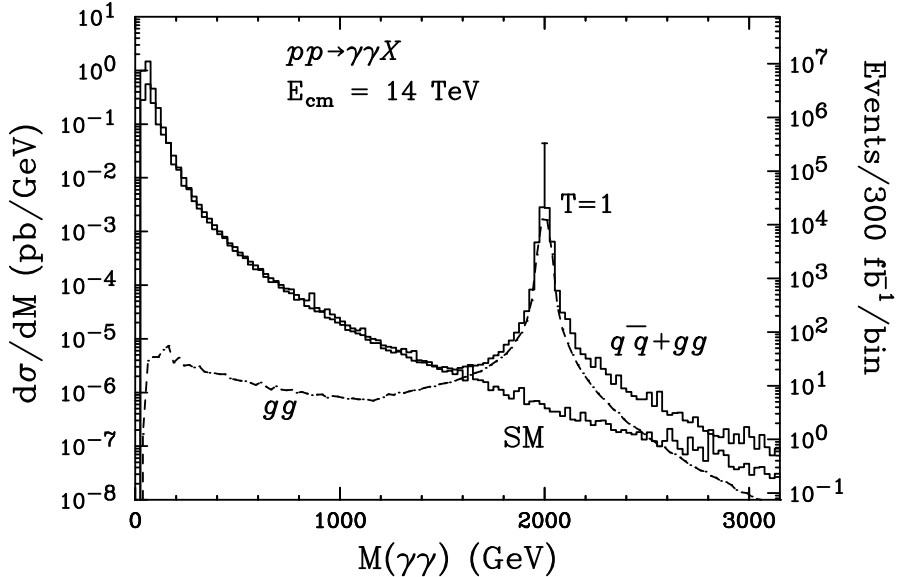}
\caption{Invariant mass distributions for  diphoton production at the LHC, for the
continuum SM expectation and the SR contributions with $\ms=2$ TeV
and $T=1$: $q\bar q+gg$ (top curve) and $gg$ only (dashed). 
The vertical bar at the $n=1$ SR peak indicates the enhancement
for $T=4$. }
\label{mgg}
\end{figure}

The differential cross-sections for diphoton production are shown in Fig.~\ref{mgg}
for the SM background and the string resonant contribution.
The diphoton processes have only odd-$n$ SRs and thus the peak is at $\ms$ for $n=1$. 
The contribution from  $gg\to \gamma\gamma$ is again separately  shown 
for comparison (dashed curve).    
 Although it would just double the diphoton signals at the peak of SR
 by including the $gg$ channel, we have pointed out earlier that the
string  coupling identification to $e$ is ambiguous. 

\subsection{Angular distributions}

\begin{table}[h]
\caption{
\label{table:SRangs}}
\tabcolsep=.5cm  
\def\arraystretch{1.5}  
\begin{tabular}{|ll|ll|}
\hline
{ process} && { angular dependence}&\\
\hline
\underline{$q\bar{q}\to \ell\bar{\ell}$}  &&& \\
$n = 1,$&$ j = 1$  &
$(d^{1}_{1,-1})^2 + (d^{1}_{1,1})^2 \propto$&$ 1+\cos^{2}\theta$  \\
        &$ j = 2$  &
$(d^{2}_{1,-1})^2 + (d^{2}_{1,1})^2 \propto$&$ 1-3\cos^{2}\theta+4\cos^{4}\theta$ \\
$n = 2,$&$ j = 1$  &
$(d^{1}_{1,-1})^2 + (d^{1}_{1,1})^2 \propto$&$ 1+\cos^{2}\theta$ \\
        &$ j = 2$  &
$(d^{2}_{1,-1})^2 + (d^{2}_{1,1})^2 \propto$&$ 1-3\cos^{2}\theta+4\cos^{4}\theta$\\
        &$ j = 3$  &
$(d^{3}_{1,-1})^2 + (d^{3}_{1,1})^2 \propto$&$ 1+111\cos^{2}\theta$\\
&&&$-305\cos^{4}\theta+225\cos^{6}\theta$\\
\hline
\underline{$gg\to \ell\bar{\ell}$}  &&& \\
$n = 1,$&$ j = 2$  &
$(d^{2}_{2,-1})^2 + (d^{2}_{2,1})^2 \propto$&$ 1-\cos^{4}\theta$  \\
 \hline \hline
\underline{$q\bar{q}\to \gamma\gamma$}  &&& \\
$n = 1,$&$ j = 2$  &
$(d^{2}_{2,-1})^2 + (d^{2}_{2,1})^2 \propto$&$ 1-\cos^{4}\theta$ \\
 \hline
\underline{$gg\to \gamma\gamma$}  &&& \\
$n = 1,$&$ j = 2$  &
$(d^{2}_{2,-2})^2 + (d^{2}_{2,2})^2 \propto$&$ 1+ 6\cos^{2}\theta + \cos^{4}\theta$  \\ 
 \hline
 
\end{tabular}
\end{table}

As already seen from Table~\ref{table:SRamps},
there are interesting mass-degeneracies with different angular
momentum states. This will lead to distinctive angular distributions 
when the pair invariant mass is close to the string resonance.
It is thus tempting to explore how this unique aspect could be studied.

\begin{figure}
\centering
\epsfxsize=4.0in
\hspace*{0in}
\epsffile{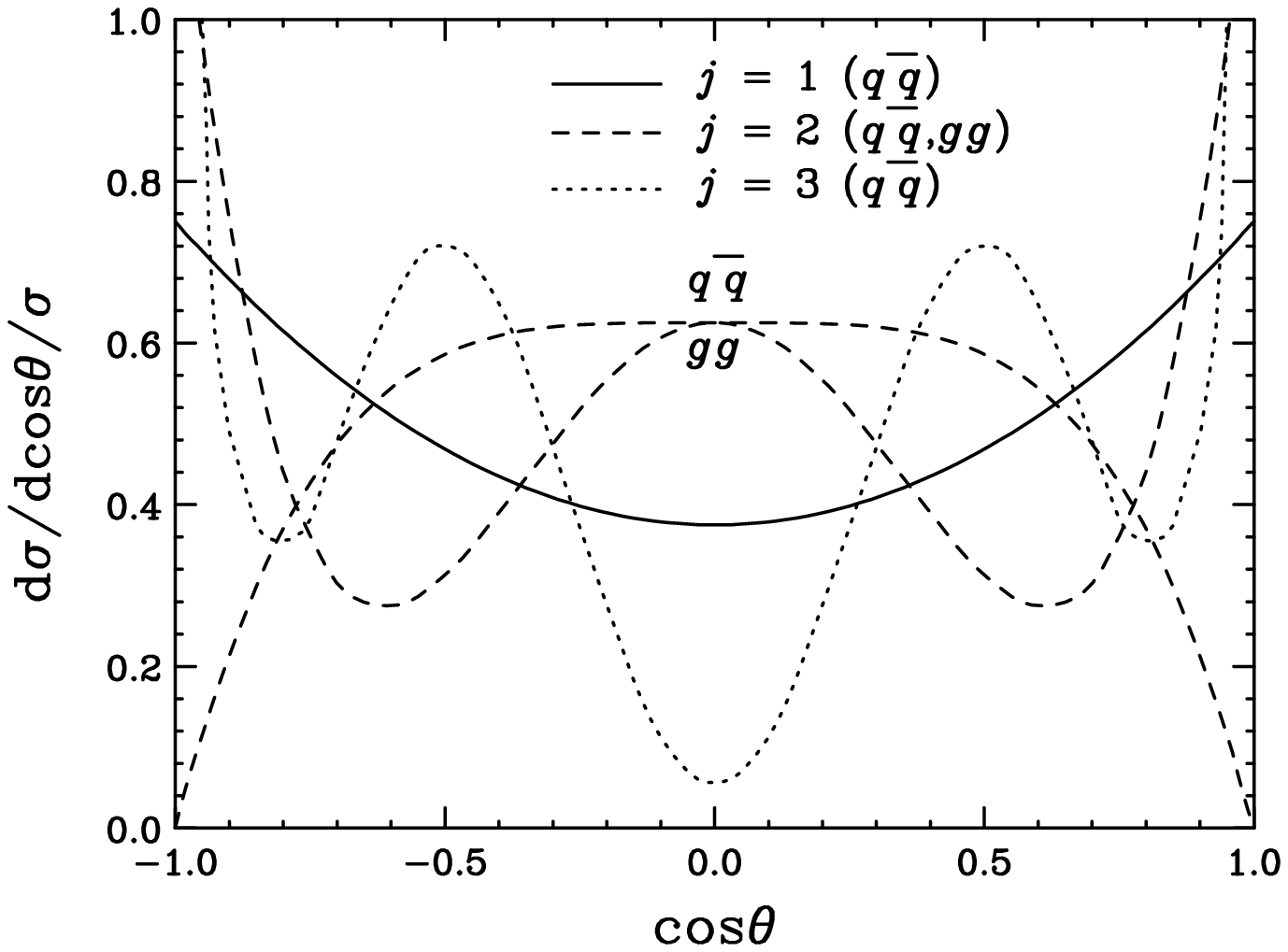}
\caption{Normalized theoretical angular distributions of string
resonances with spin 1, 2, and 3 in the DY channel $pp\to \ell^+\ell^-X$. }
\label{cth}
\end{figure}

\begin{figure}
\centering
\epsfxsize=4.0in
\hspace*{0in}
\epsffile{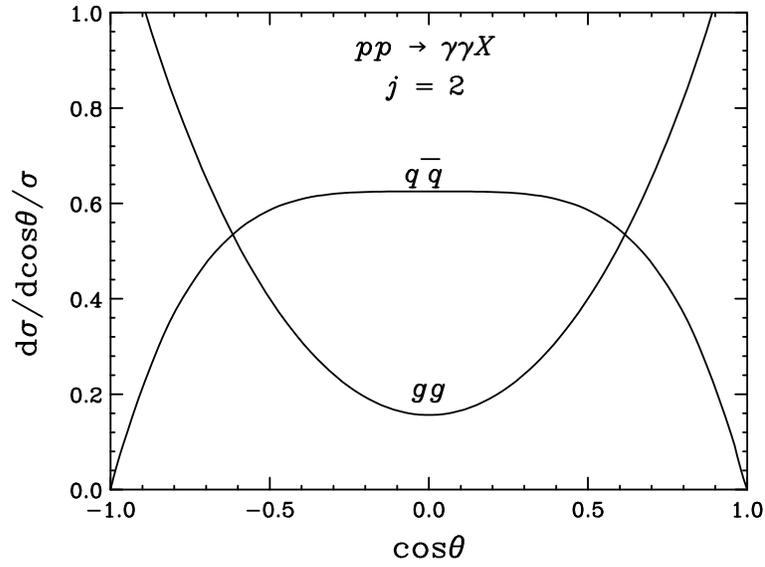}
\caption{Normalized theoretical angular distributions of string
resonances with only spin-2 in $pp\to \gamma\gamma X$. }
\label{cgg}
\end{figure}

We first tabulate the angular dependence for the processes with given
$n,\ j$ values in Table~\ref{table:SRangs}.  As always, the angle 
$\theta$ is defined in the $\ell\bar \ell$ or $\gamma\gamma$ rest frame
with respect to the beam direction.  It is indeed interesting to
see the drastic differences of the angular distributions for different
processes.  For instance, there is a degeneracy of spin 1 and 2 at the first SR in dileptonic processes.  Spin-2 contributions to dileptonic processes have two possible sources with totally different angular distributions.  One is from SR of $q\bar{q}$ initial state and another is from SR of $gg$ one as illustrated in Fig.~\ref{cth} by the dashed curves. Here,   
the contribution of spin-2 SR from $q\bar{q}$ is 
one-ninth of the spin-1 contribution of the same process while the contribution from $gg$ is 
directly proportional to the Chan-Paton parameter $T$.  
These two contributions of spin-2 exchange could change the angular distribution significantly from the conventional ``$Z^{\prime}$" exchange that we would encounter in many extensions of
the SM \cite{zp,lan,carena}.
It is obvious that this unique angular distribution is also distinguishable from new-physics models with only spin-2 exchange such as Kaluza-Klein graviton \cite{aopw}.  For diphoton processes, there  
is only spin-2 SR from both $q\bar{q}$ and $gg$ initial states, as shown in Fig.~\ref{cgg}.  

\begin{figure}[tb]
\centering
\epsfxsize=4.0in
\hspace*{0in}
\epsffile{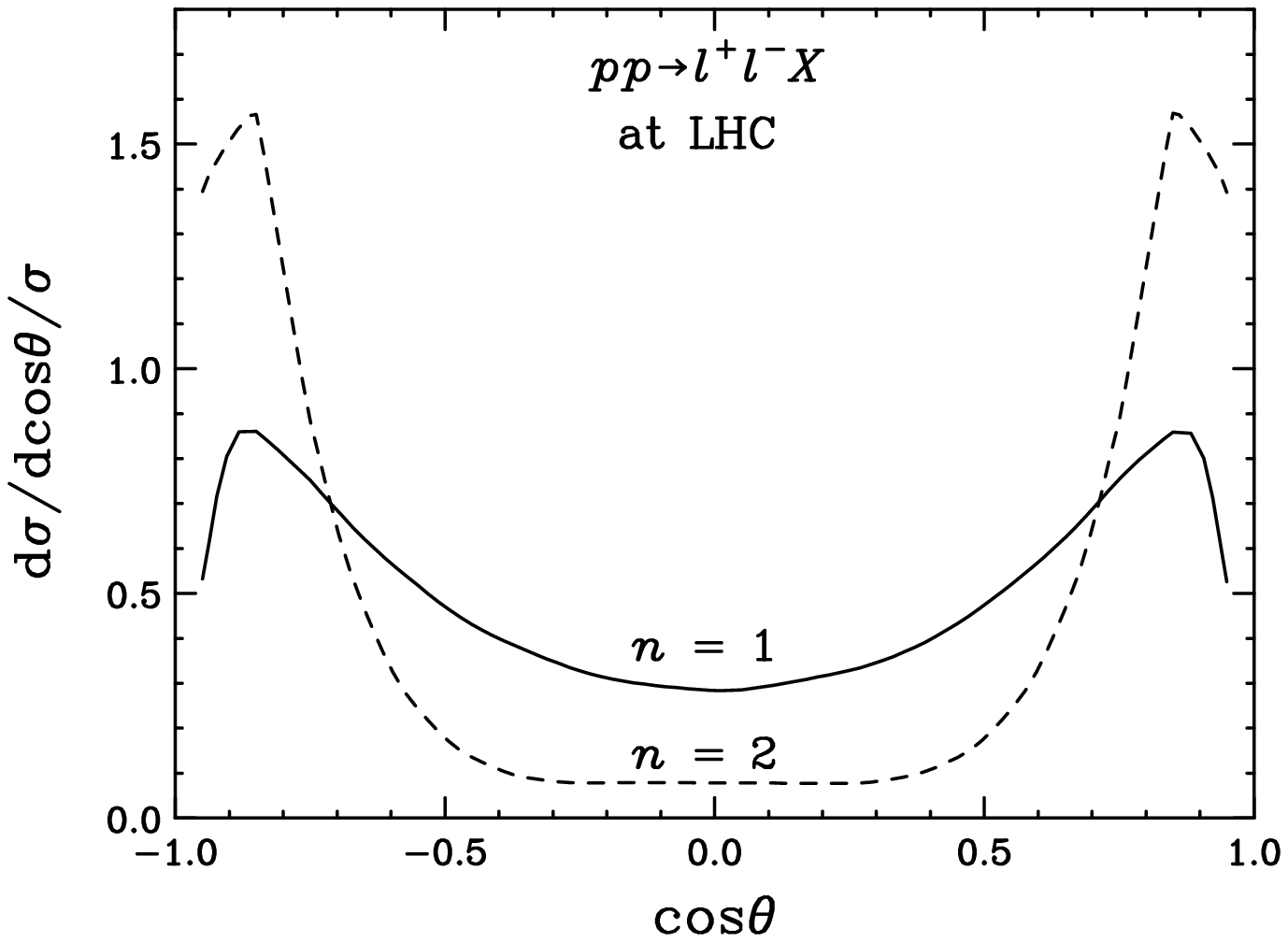}
\caption{Normalized angular distributions for $n=1$ (solid) and $n=2$ (dashed) string
resonances in the DY channel $pp\to \ell^+\ell^-X$ with appropriate cuts
of Eq.~(\ref{cuts}). }
\label{c-ll}
\end{figure}

\begin{figure}[tb]
\centering
\epsfxsize=4.0in
\hspace*{0in}
\epsffile{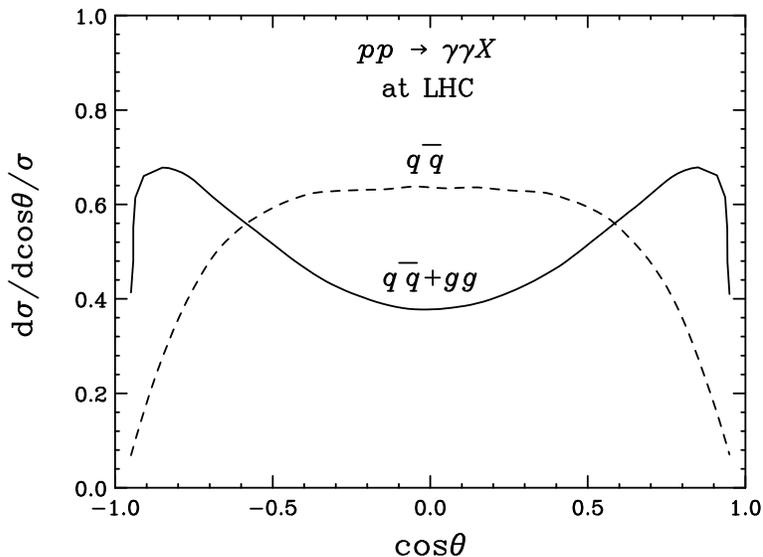}
\caption{Normalized angular distributions for $n=1$ string resonance 
in the diphoton channel $pp\to \gamma\gamma X$ with appropriate cuts
of Eq.~(\ref{cuts}). The solid curve represents
the total contribution of $q\bar{q}+gg$ and the dashed curve is for $q\bar q$ only. }
\label{c-gg}
\end{figure}

In Figure~\ref{c-ll}, the predicted angular distributions (normalized to unity)  
of dileptonic signals  are presented with the choice of $T=1$ for both $q\bar{q}$ and $gg$ initial states,
for two different mass eigenstates $n=1,2$. The events are selected not only by imposing the
acceptance cuts of Eq.~(\ref{cuts}), but also by choosing the invariant mass around the
resonance mass
\begin{equation}
\sqrt{n}M_S-2\Gamma_n< M < \sqrt{n}M_S+2\Gamma_n.
\end{equation}
We see from the figure that the distribution for $n=1$ is less pronounced near $\cos\theta\sim \pm1$
than that for $n=2$. The eventual drop is due to the acceptance cuts. One could imagine to
fit the observed distributions in Fig.~\ref{c-ll} by the combination of the functions listed in
Table~\ref{table:SRangs} to test the model prediction.
Similar distribution for the $\gamma\gamma$ final state is shown in Fig.~\ref{c-gg},
where the total contribution of $q\bar{q}+gg$ (the solid curve)
and that for $q\bar q$ only (the dashed curve) are compared at $T=1$ for both processes.

\subsection{The Forward-Backward asymmetry}

For parton-level subprocess $q\bar{q}\to \ell\bar{\ell}$, forward-backward asymmetry is defined as 
\begin{eqnarray}
A^{q\ell}_{FB} & = & \frac{N_{F}-N_{B}}{N_{F}+N_{B}}
\end{eqnarray}
where $N_{F(B)}$ is the number of events with final lepton moving into the forward (backward) direction.  At $pp$ colliders, the annihilation process is from the valence quarks and the sea antiquarks.  Therefore, the produced intermediate resonant state will most likely move along the direction of the initial valence quark due to its higher fraction of momentum \cite{lan}.  With respect to one particular boost direction of the final dilepton, we can consequently extract information of the forward-backward asymmetry of the subprocess.

In our open-string model, the asymmetry is given, for $s\gg m^2_Z$, by
\begin{eqnarray}
A^{q\ell}_{FB} & = & \left(\frac{30}{32}\right)\frac{(G^{q}_{LL})^2+(G^{q}_{RR})^2-(G^{q}_{LR})^2-(G^{q}_{RL})^2}{(G^{q}_{LL})^2+(G^{q}_{RR})^2+(G^{q}_{LR})^2+(G^{q}_{RL})^2}\\
\nonumber \\ 
               & = & \left\{ \begin{array}{ll}  -0.176 ~(-0.039) & \mbox{      for $q=u, T=1 ~(4)$} \\
\\
                                           0.160 ~(0.042) & \mbox{      for $q=d, T=1 ~(4)$}
\end{array}
\right.
\end{eqnarray}
where $G^{q}_{\alpha\beta}=F_{\alpha\beta}+2T$, the interaction factor of the fermions 
defined in Sec.~\ref{ampl}.  This asymmetry is inherited from the SM part, $F_{\alpha\beta}$, in the amplitudes.  The value of $A^{q\ell}_{FB}$ for SM with $s\gg m^2_Z$
is $0.61~(0.69)$ for $u~(d)$ quark.
The asymmetry is diluted by the symmetric SR contribution 
 since typically $T>F_{\alpha\beta}$.  
The forward-backward asymmetry is hardly visible when $T=4$.
This also can be viewed as another feature to distinguish the SR from the other
states like $Z'$ which normally yields larger asymmetry \cite{lan}.

\subsection{The reach on the string scale}

\begin{figure}
\centering
\epsfxsize=4.0in
\hspace*{0in}
\epsffile{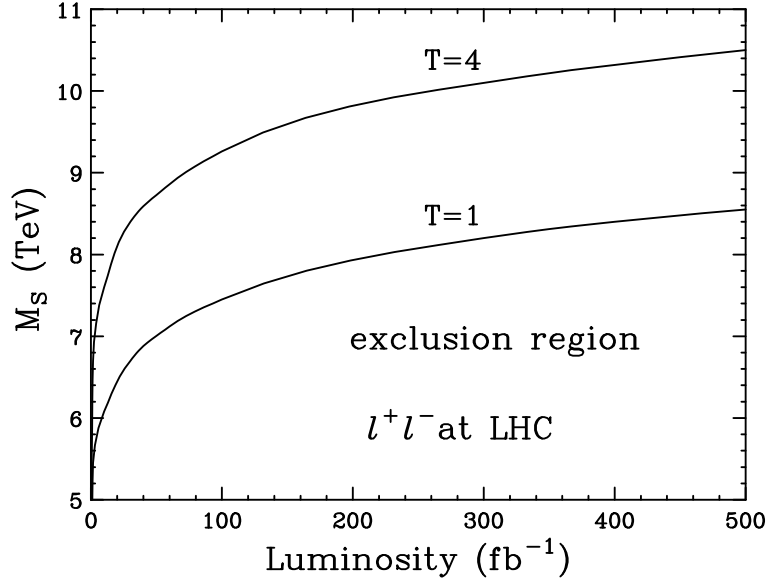}
\caption{Sensitivity reach at $95\%$ C.L. of $M_S$ at various luminosities at the LHC.}
\label{sens}
\end{figure}

For the unfortunate possibility that we do not detect any signals with SR properties, the absence of signals implies certain bound on the string scale  $M_S$ and Chan-Paton parameters $T$. 
We present the sensitivity reach at $95\%$ C.L.~in Fig.~\ref{sens} 
as a function of the integrated luminosity at the LHC. The results are obtained by assuming 
the  Gaussian statistics and by demanding $S/\sqrt{S+B}>3$, where the signal rate is estimated
in the dilepton-mass window $[M_S-2\Gamma_{1}, M_S+2\Gamma_{1}]$ at the first SR.  
The lower bound on the string scale could reach 
$M_S>8.2-10$ TeV for $T=1-4$ at a luminosity of 300 fb$^{-1}$.

\section{Summary and Conclusions}
\label{conclusions-sec}
We have constructed tree-level open-string amplitudes for dilepton and diphoton processes.
The massless SM particles are identified as the stringy zero-modes.
For a given $2\to 2$ scattering process, by demanding the open-string amplitudes 
reproduce the SM ones at low energies, the amplitudes can be casted into a generic form 
\begin{equation}
A_{string} \sim A_{SM}(s,t,u)\cdot S(s,t, u) + T f(s,t,u)\cdot g(s,t,u),
\label{form}
\end{equation}
where $A_{SM}$ is the SM amplitude, $S(s,t,u) = S(s,t),S(s, u)$ or $S(t,u)$ the Veneziano amplitudes,
$T$ the undetermined Chan-Paton parameter, $ f(s,t,u)$ a kinematical function
given in Eq.~(\ref{eq:f}), and $g(s,t,u)$ some process-dependent kinematical function. The amplitudes have the following general features:
\begin{itemize}
\item By construction, they reproduce the standard model amplitudes 
at low energies $s\ll \mssq$, since $S(s,t)\to 1$ and $f(s,t,u)\to 0$,  
and thus fixing the string couplings with respect to the SM gauge couplings.
\item  The Veneziano amplitude $S(s,t)$ and $f(s,t,u)$ 
develop  stringy resonances at energies $\sqrt s=\sqrt n \ms\  (n=1,2,...)$.
\item $S(s,t)$ leads to both even- and odd-$n$ resonances, while $f(s,t,u)$ yields 
 only odd-$n$ SRs. 
 Thus, the even-$n$ resonances are completely fixed by the SM interactions,
 independent of the unknown factor $T$.
\item 
For the standard model processes that either vanish at tree-level 
(such as $g g \to \gamma\gamma$), or do not contain 
$s$-channel exchange (such as $q\bar{q}\to \gamma\gamma$), 
there will be no SRs which couple with SM charges as in the first term of
Eq.~(\ref{form}).  Yet, there can still be SR contributions from purely stringy
effects, directly proportional to $T$,  given in the second term of the equation.
\end{itemize}

We would like to emphasize the profound implication of our amplitude construction
and the generic structure of Eq.~(\ref{form}). 
The basic assumption of this work is to take the tree-level
open-string scattering amplitudes of Eq.~(\ref{eq:1})
as the description of leading new physics beyond the SM near the TeV threshold.
As long as one accepts this approach and demands the amplitudes 
to reproduce the SM counterparts at low energies, 
Eq.~(\ref{form}) would be the natural consequence. There are essentially
only two unknown parameters: the string scale $\ms$ and the Chan-Paton
parameter $T$. This construction should be generic for  any leading-order
$2\to 2$ processes of massless SM particle scattering, and thus be applicable 
for further phenomenological studies.

We have calculated numerically the total cross-section of DY through the first string resonance and compared with the CDF data for $Z^{\prime}$ production.  We establish the current lower bound 
of the string scale at about $1.1-2.1$ TeV which is stronger than limits from the contact-interaction analysis \cite{bhhm}.  The bound from Tevatron can be improved to $1.5-3$ TeV with an integrated
luminosity of 2 fb$^{-1}$.

At the CERN LHC, with the high luminosity expected and much 
larger center-of-mass energy, SR-induced signals for $M_S \lsim 8$ TeV 
can be substantial and a large number of events is predicted around the SR  
in dilepton and diphoton  processes regardless of the value of the Chan-Paton 
parameters $T$.  The second string resonance with a mass $\sqrt 2 \ms$
may be observed in the dilepton channel as well. 
Distinctive angular distributions and the forward-backward asymmetry
may serve as indicators to distinguish the SR from other new physics. 
For a larger value of $M_S$, 
SR signals become weaker and we may establish the sensitivity on the 
lower bound of the string scale for $T=1-4$ to be $M_S>8.2-10$ 
TeV at $95\%$ C.L. with a luminosity of 300 fb$^{-1}$.        

\section*{Acknowledgments}
\indent
We would like to thank Doug McKay for helpful comments and discussions. 
  This work was supported in part by the U.S. Department of Energy under 
  grant number DE-FG02-95ER40896,  and in part by the Wisconsin
Alumni Research Foundation.
T.H. was also supported in part  by a Fermilab Frontier Fellowship, and 
by the National Natural Science Foundation of China.
 Fermilab is operated by the Universities Research Association 
 Inc.~under Contract No.~DE-AC02-76CH03000 with the U.S.~Department of Energy.

\appendix

\section{kinematic table}

Consider a tree-level scattering of four massless gauge bosons 
in $SU(N)$ gauge theory, with  all momenta incoming. The only non-vanishing
amplitudes are those with two positive and two negative helicities. There are
six of them, each as a sum of three terms of independent permutations. 
The general formula for one permutation is given in Ref.~\cite{man} as
\begin{eqnarray}
A_{1234} = ig^2 \frac{\langle IJ \rangle^4}{\langle12\rangle
\langle23\rangle \langle34\rangle \langle41\rangle},
\label{eq:a}
\end{eqnarray}
where $I, J$ label the two gauge bosons with negative helicities.  
Obviously, the above amplitude is invariant if $I,J$ are for the positive
helicity gauge bosons.
$\langle pq \rangle$ is the spinor product defined by
\begin{eqnarray}
\langle pq \rangle \equiv \overline{\Psi_{-}(p)}\Psi_{+}(q)
\end{eqnarray}
and $|\langle pq \rangle|^2 = 2 p\cdot q$.  The order of $\langle XY \rangle$ 
in the denominator is cyclic of 1234.  For processes involving fermions, 
the supersymmetric relation of Eq.~(4.9) in \cite{man} can been applied.  
The expressions for four fermions ($ffff$)  are exactly the same as those 
for four gauge bosons ($gggg$) for each corresponding helicity and particle permutation.  The amplitudes for processes with two bosons and two fermions vanish when the 
 two fermions (or bosons) have the same helicity. 
A useful list of the amplitudes relevant to our scattering amplitude  construction in the text 
is   given as follows, where the superscripts indicate the helicities with respect to the incoming 
momenta. 

%

{\tabcolsep=.3cm  
\def\arraystretch{2.5}  
\def\dis{\displaystyle}  
\begin{tabular}{llll}
$g^\pm g^\mp g^\mp g^\pm/f^\pm f^\mp f^\mp f^\pm:$ 
&  $A_{1234}=ig^2 \frac{\langle14\rangle^2}{\langle12\rangle^2}$ 
                    &      $A_{1324} = ig^2 \frac{\langle14\rangle^2}{\langle13\rangle^2}$ 
                    &      $A_{1243} = ig^2 \frac{\langle14\rangle^4}{\langle12\rangle^2 \langle13\rangle^2}$
 \\[.5cm]
$g^\pm g^\mp g^\pm g^\mp/f^\pm f^\mp f^\pm f^\mp:$ 
 & $A_{1234} = ig^2 \frac{\langle13\rangle^4}{\langle12\rangle^2 
\langle14\rangle^2} $
                                & $A_{1324} = ig^2 \frac{\langle13\rangle^2}{\langle14\rangle^2}$
                                & $A_{1243} = ig^2 \frac{\langle13\rangle^2}{\langle12\rangle^2}$ 
 \\[.5cm]
$g^\pm g^\mp f^\mp f^\pm/ f^\mp f^\pm g^\pm g^\mp:$ 
&  $A_{1234}=ig^2 \frac{\langle13\rangle \langle14\rangle}{\langle12\rangle^2}$
                    &       $A_{1324} = ig^2 \frac{\langle14\rangle}{\langle13\rangle}$
                    &       $A_{1243} = ig^2 \frac{\langle14\rangle^3}{\langle13\rangle \langle12\rangle^2}$ 
 \\[.5cm]
\end{tabular} }

 Expressions  for other helicity combinations can be achieved 
 by  properly crossing two particle momenta, or by cyclic permutation under
 which Eq.~(\ref{eq:a}) is invariant.
In doing so, some identities may be useful: 
\begin{itemize}
\item $A_{ijkl} = A_{lkji};\quad  A_{ijkl} = A_{ilkj}$; 
\item invariant under the  sign change  $(++\leftrightarrow --)$.
\end{itemize}

\section{calculation of decay widths}
 
The partial decay width of SR with a mass $m=\sqrt n \ms$ and angular momentum $j$
to a final state $\ell\bar \ell$ can be written generically as
\begin{eqnarray}
\Gamma^j_{n} & = & \frac{1}{2m}\frac{1}{2j+1}\int dPS_2 |A(X^j_{n} \rightarrow \ell\bar{\ell})|^2.
\end{eqnarray}
The two-body phase space element is $dPS_2=d\Omega/8$, and the decay matrix element
squared can be related to the scattering amplitude by 
\begin{equation}
|A(X^j_{n} \rightarrow \ell_3\bar{\ell_4})|^2=(s-m^2)
|A^j_{n}(\ell_1\bar{\ell_2} \rightarrow \ell_3\bar{\ell_4})|\quad {\rm with}\ p_{1}=p_{3},\ p_{2}=p_{4}.
\end{equation}
With the help of partial wave expansion in terms of the Wigner functions 
$d^j_{mm^{\prime}}$ as discussed in Sec.~\ref{part}, we have
\begin{eqnarray} 
A^j_{n}(\ell_{\alpha}\bar{\ell}_{\beta} \rightarrow \ell_{\alpha}\bar{\ell}_{\beta}) 
& = &ig^2 G_{\alpha\alpha} {s\ \alpha^j_{n}\ d^j_{1,-1}\over s-m^2 }.
\label{expand}
\end{eqnarray}
where
\begin{equation}
\nonumber
G = \left\{
\begin{array}{l}
 F + 2T \quad {\rm for\ odd}\  n , \\[3mm]
\nonumber
 F~~~~ \qquad {\rm for\  even}\  n ,
\end{array}
\right.
\end{equation}
with $F$ and $T$ given in text. 
The coefficient $\alpha^j_{n}$ satisfies normalization condition $\sum_{j=1}^{n+1}|\alpha^j_{n}|=1$.  
The final expression for decay width of the SR is therefore
\begin{eqnarray}
\Gamma^j_{n}  = \frac{g^2}{16\pi}\ \frac{\sqrt{n} M_S}{2j+1} \ 
G_{\alpha\alpha}\ | \alpha^{j}_n |
\end{eqnarray}  
This expression can be easily generalized to other elastic processes. 
As for  the case of diphoton production,
the gauge coupling factor $G=T$ after absorbing the $1/2$ factor for identical particles, 
and the coupling $g^2/16\pi = \alpha/4$, instead of $\alpha/4\xw$ as in the dilepton case.  
It should also be noted that even 
we do have a non-vanishing SM part in the $q\bar{q}\gamma\gamma$ channel, 
there is no corresponding contribution from an SR and consequently 
to the width of diphoton processes.  

\begin{table}[tb]
{\tabcolsep=.5cm  
\def\arraystretch{1.5}  
\medskip
\centering
\begin{tabular}{|cc|ccc|} \hline
&$n $&  $j=1$  &  $2$  &  $3$  \\ \hline
$q\bar{q}\ell\bar{\ell}$  & $1$ & $3/4$  &  $\mp 1/4$ & $0$  \\
$                      $  & $2$ & $-9/20$  &  $\pm 5/12$ & $-2/15$  \\ \hline
$q\bar{q}\gamma\gamma$  & $1$ & $0$  &  $-1$ & $0$  \\
$                    $  & $2$ & $0$  &  $0$ & $0$  \\ \hline
$gg\ell\bar{\ell}$  & $1$ & $0$  &  $-1$ & $0$  \\
$                $  & $2$ & $0$  &  $0$ & $0$  \\ \hline
$gg\gamma\gamma$  & $1$ & $0$  &  $1$ & $0$ \\
$              $  & $2$ & $0$  &  $0$ & $0$  \\ \hline
\end{tabular} }
\caption[]
{\label{alpha}
\small
Coefficients $\alpha^{j}_n$ of partial wave expansion in each processes.  
Upper (lower) sign in $q\bar{q}\ell\bar{\ell}$ corresponds to  scattering of 
quark into lepton with like (opposite) helicity.} 
\end{table}

For completeness,  in Table \ref{alpha} 
we provide the expansion coefficients in Eq.~(\ref{expand}),
and the relevant Wigner functions are
\begin{eqnarray}
d^1_{1,-1}& = & \frac{1-\cos\theta}{2}\\ 
d^2_{1,-1}& = & \frac{1-\cos\theta}{2}(2\cos{\theta}+1)\\ 
d^3_{1,-1}& = & \frac{1}{4}\left(\frac{1-\cos\theta}{2}\right)(15\cos^2{\theta}+10\cos{\theta}-1)\\ 
d^2_{2,-1}& = & -\sin{\theta}\left(\frac{1-\cos\theta}{2}\right)\\ 
d^2_{2,-2}& = & \left(\frac{1-\cos\theta}{2}\right)^2 
\end{eqnarray}
with $d^j_{1,1}(x)=(-1)^{j-1}d^j_{1,-1}(-x)$ and $d^2_{2,m}(x)=d^2_{2,-m}(-x)(m= 1, 2)$.

Numerically, the total widths for each processes when $T=1$ are
\begin{eqnarray}
\Gamma^{1,2}_1(q\bar{q}\ell\bar{\ell})& = & 240, ~48 \mbox{ GeV}
\left(\frac{M_S}{\mbox{TeV}}\right),\\
\Gamma^{1,2,3}_2(q\bar{q}\ell\bar{\ell})& = & 46, ~26, ~5.8   \mbox{ GeV}
\left(\frac{M_S}{\mbox{TeV}}\right),\\
\Gamma^{2}_1(gg\ell\bar{\ell})& = & 19  \mbox{ GeV}
\left(\frac{M_S}{\mbox{TeV}}\right),\\
\Gamma^{2}_1(q\bar{q}\gamma\gamma)  & = & 3.9 \mbox{ GeV}
\left(\frac{M_S}{\mbox{TeV}}\right),\\
\Gamma^{2}_1(gg\gamma\gamma)  & = & 3.5 \mbox{ GeV}
\left(\frac{M_S}{\mbox{TeV}}\right).
\end{eqnarray}
where we have included all necessary 
decay modes into related final states for each resonance.  
For instance, the width $\Gamma(q\bar{q}\ell\bar{\ell})$  includes 
the partial decay widths of SR into charged leptons, neutrinos, and quarks.
Partial decay modes into massive bosons such as the Higgs and $W^\pm, Z$ are not included.

\end{document}